\documentclass[aps,prl,twocolumn,groupedaddress,amsmath,amssymb]{revtex4}
\usepackage{bbm}
\usepackage{mathrsfs}
\usepackage{amsfonts}
\usepackage{color}
\usepackage{makecell}
\def\newblock{\hskip .11em plus .33em minus .07em}

\newcommand{\sv}{{\boldsymbol s}}

\newcommand{\Tr}{{\rm Tr}}

\newcommand{\be}{\begin{equation}}
\newcommand{\ee}{\end{equation}}
\newcommand{\ba}{\begin{eqnarray}}
\newcommand{\ea}{\end{eqnarray}}
\begin{document}
\title{Action Principle and Dynamic Ensemble Theory for Non-equilibrium Markov Chains}
\author{Xiangjun Xing$^{1,2}$}
\email{xxing@sjtu.edu.cn}
\author{Mingnan Ding$^{1}$}
\email{dmnphy@sina.com}
\address{$^1$Wilczek Quantum Center, School of Physics and Astronomy, Shanghai Jiao Tong University, Shanghai, 200240 China, 
$^2$Collaborative Innovation Center of Advanced Microstructures, Nanjing 210093, China}

\date{\today} 


\begin{abstract}
An overarching action principle, {\em the principle of minimal free action}, exists for ergodic Markov chain dynamics.  Using this principle and the Detailed Fluctuation Theorem, we construct a dynamic ensemble theory for non-equilibrium steady states (NESS) of Markov chains, which is in full analogy with equilibrium canonical ensemble theory.  Concepts such as energy, free energy, Boltzmann macro-sates,  entropy, and thermodynamic limit all have their dynamic counterparts.  For reversible Markov chains, minimization of {\em Boltzmann free action}  yields thermal equilibrium states, and hence provide a dynamic justification of the {\em principle of minimal free energy}.   For irreversible Markov chains, minimization of {\em Boltzmann free action} selects the stable NESS, and determines its macroscopic properties, including entropy production.  A quadratic approximation of  free action leads to linear-response theory with reciprocal relations built-in. Hence, in so much as non-equilibrium phenomena can be modeled as Markov processes, minimal free action serves as a basic principle for both equilibrium and non-equilibrium statistical physics.  


\end{abstract}
\maketitle


\paragraph{Introduction.} It is remarkable that variational principles~\cite{action-principle-scholarpedia} underly many basic laws of physics, including those in Hamiltonian mechanics, geometric optics, quantum mechanics and equilibrium statistical mechanics.   For example, the entire theory of equilibrium statistical mechanics can be formulated on the base of the principle of maximal entropy.   Any equilibrium phenomenon can be claimed understood as soon as we can derive it from this principle.  

Search for a similar unifying action principle of non-equilibrium phenomena has been a long and inconclusive quest.  The earliest versions of   ``minimum dissipation theorem'' were stated by Helmholtz in 1869~\cite{Helmholtz-1869}, and by Rayleigh in 1873~\cite{Rayleigh-1873}. It was generalized substantially by Onsager~\cite{Onsager-1931,Onsager-Machlup-1953,Gyarmati-1970}, such that it serves a starting point for linear response theory.  Another candidate is ``the principle of minimal entropy production'', studied by Prigogine~\cite{Min-EP-Prigogine} and many others~\cite{Klein-1954,Jaynes-1980,Nicolis-1984,Maes-2007,Schnakenberg-review}, and  caused major confusions and debates.  For a more recent exposition see reference \cite{Min-EP-scholarpedia}.  There are also many other proposals~\footnote{ A detailed discussion of history can be found in the wikipedia article titled {\it Extremal principles in non-equilibrium thermodynamics}~\cite{wiki-extremal-principle}.  }, including the principle of  ``maximal entropy production''~\cite{Paltridge-max-disispation,Max-EP-review}, which is in apparent contradiction with the the previous two principles. The relations between these proposed principles have been are addressed repeatedly, but not yet completely understood.  We shall not elaborate on these issues  as our approach will be substantially different.  


Recent progresses in study of stochastic thermodynamics~\cite{Jarzynski-review,Seifert-review,Sekimoto-book,Evans-Searles,Pitaevskii-2011} have supplied much fresh understanding of entropy production and dissipation in non-equilibrium processes.  It has become clear that entropy production can be defined at the level of individual dynamic path, and that it changes sign under time-reversal.  Given a non-equilibrium boundary conditions or driving forces, the entropy production is positive for some dynamic paths and negative for their time reversal counterparts.  Fluctuation Theorems~\cite{Crooks-FT-1,Gallavotti-Cohen-1995,Lebowitz-Spohn-1999,Kurchan-1998} tell us that those paths with positive dissipation are exponentially more probable than their time-reversal, which have negative dissipation.  A principle of ``minimal entropy production'' or of ``least dissipation'', if taken literally, would select those paths with negative entropy production, and therefore constitutes an outright contradiction to the second law of thermodynamics.  

Most, if not all, non-equilibrium problems can be modeled as Markov processes.   Inspired by the conceptual progresses in stochastic thermodynamics, we feel that the theory of ergodic Markov processes can be formulated as a dynamic ensemble theory.  This would allow us to use all powerful tools in equilibrium theory to study non-equilibrium physics, and hence greatly facilitate theoretical understanding of non-equilibrium statistical physics.   This is indeed the case, as we will demonstrate below.   As enumerated in Table \ref{table:rsd}, many important concepts of equilibrium statistical mechanics, including micro-state, macro-state, energy, entropy, free energy all have their dynamic counterparts in Markov processes.  The only exception is temperature, whose dynamic counterpart is unity.  Most importantly, we will find that there is also an overarching action principle, i.e. {\em the principle of minimal free action}, which serves as a foundation of dynamic ensemble theory for Markov chains.  This principle governs all macroscopic properties of non-equilibrium steady states, including the equations of state and conditions of stability.  At the quadratic level, the principle yields linear response theory with reciprocal symmetry naturally built-in.  For systems with time-reversal symmetry, the principle of minimal free action selects the thermal equilibrium state with minimal free energy, and hence constitutes a dynamic justification of the principle of maximal entropy.  Consequently, under the reasonable assumption that non-equilibrium phenomena can be modeled as Markov processes, minimal free action serves as a basic principle for both equilibrium and non-equilibrium statistical physics.    

\begin{center}
\begin{table}[tbh!]
\begin{ruledtabular}
\setlength{\extrarowheight}{4pt}
\begin{tabular}{c | c }
\makecell{{\bf Equilibirium} \\ {\bf Ensemble Theory} } 
&    \makecell{{\bf  Dynamic} \\  {\bf Ensemble Theory} } 
 \\ \hline\hline
Micro-states &  Dynamic paths   \\\hline
Number of Spins &   Number of Time Steps \\\hline
Thermodynamic Limit & Long Observation Time \\\hline
\makecell{ Energy  $E({\mathbf s})$ }
	& Action ${\mathcal E}(\ell)$ \\\hline
Mag. Energy $ - B M({\mathbf s})$ & Entropy Prod. $-\Sigma(\ell)/2$ \\\hline
Temperature $T$ & 1 \\\hline
Prob. of Micro-state $\rho(\sv)$ &  Prob. of Path $q(\ell)$ \\\hline
Gibbs Entropy $S_G[\rho]$ &  Dynamic Entropy $H[q]$ \\\hline
Free Energy $F[\rho]$ &  Free Action ${\mathcal F}[q]$ \\\hline
\makecell{ Boltzmann Entropy \\ $S_B(E,M)$} & \makecell{Dynamic Boltz. Entropy \\ $H_B(\Theta, \Sigma) $ }\\\hline
Boltz. Free Energy $F_B$ & Boltz. Free Action $\Phi_B$ \\\hline
 \makecell{\bf Principle of \\\bf Minimal Free Energy } & 
\makecell{\bf Principle of \\  \bf Minimal Free Action} \\\hline
\makecell{ Equations of State \\$\delta^1 F_B = 0 $} & 
\makecell{  Equations of State \\ $ \delta^1 \Phi_B = 0$} \\\hline
Stability Cond. $\delta^2 S_B \geq 0$ & Stability Cond.  $\delta^2 H_B \geq 0$ \\\hline
Symmetry of Correlations & Onsager's Reciprocal Relations
\end{tabular}
\end{ruledtabular}
\caption{Analogy between equilibrium and dynamic ensemble theories.  }
\label{table:rsd}
\vspace{-4mm}
\end{table} 
\end{center}

We will develop the dynamic ensemble theory in parallel to equilibrium ensemble theory.  To set the stage for comparison between equilibrium and non-equilibrium theories, we first summarize the key elements of equilibrium canonical ensemble theory and the principle of minimal free energy, using Ising model for illustration.   The detailed comparison of equilibrium and non-equilibrium ensemble theories are shown in Table~\ref{table:rsd}.  

\vspace{3mm} 
\paragraph{Sketch of equilibrium ensemble theory. }  
The micro-states of system are designated by a set of microscopic variables called spins,  $\sv = \{\sv_i\}$.  The  Hamiltonian ${\mathcal H}(\sv) $ is given by
\ba
{\mathcal H}(\sv) &=& {\mathcal H}_0(\sv)  + {\mathcal H}_1(\sv)  = {\mathcal H}_0(\sv) - B \sum \sv_i,
\label{static-H-def}
\ea
where ${\mathcal H}_0(\sv)$, called the {\em intrinsic energy}, is interaction between neighboring spins, and ${\mathcal H}_1(\sv) = - B \sum \sv_i$ is due to  external field. ${\mathcal H}_0(\sv)$  and ${\mathcal H}_1(\sv)$ are respectively even and odd under flipping of spins $\sv \rightarrow - \sv$.

The canonical ensemble theory of equilibrium statistical mechanics can be constructed from {\em the principle of minimal free energy}, which is known to be equivalent to {\em the principle of maximal entropy}.  Let $\rho (\sv)$ be an arbitrary  probability distribution in the space of micro-states. The non-equilibrium free energy $F[\rho]$ as a functional of $\rho({\sv})$ is defined as
\be
F[\rho] \equiv  \langle {\mathcal H} \rangle_\rho - T\, S[\rho]
= \Tr_{\sv} \, \rho(\sv) \left( {\mathcal H}(\sv) + T \log \rho(\sv)\right),
\label{free-energy-def}
\ee
where $\Tr_{\sv}$ means summation over all spin configurations, and  $T = \beta^{-1}$ is the temperature.  Here we  use the convention  $k_B =  1$ throughout.  Minimization of $F[\rho]$ with respect to $\rho (\sv)$ leads to the equilibrium Gibbs-Boltzmann distribution: 
\be
\rho_{\rm EQ}(\sv) = e^{-\beta {\mathcal H}(\sv)}/Z,
\label{rho-eq-1}
\ee
 where $Z = e^{-\beta F_{\rm EQ}}  = \Tr_{\sv} \, e^{-\beta {\mathcal H}}$ is the canonical partition function, and  $F_{\rm EQ} = - T \log Z$ is the equilibrium canonical free energy.  

For fixed values of intrinsic energy ${\mathcal H}_0(\sv)  = E_0$ (excluding magnetic field energy), and magnetization $\sum {\bf s} = M$, the totality of all micro-states is defined as a {\em Boltzmann macro-state} $\Omega(E_0, M)$.  Its entropy is called the Boltzmann entropy $S_B(E_0, M)$.  Mathematically we have
\vspace{-5mm}
\begin{subequations}
\label{Boltzmann-EQ-1}
\ba
\Omega(E_0, M) &=& \{\sv | {\mathcal H}_0(\sv) = E_0, \sum {\bf s} = M\}, \\
\label{Boltzmann-def-static}
S_B(E_0, M) &=& \log |\Omega(E_0, M)|, 
\label{Boltzmann-entropy-def-static}
\ea 
\end{subequations}
where $|\Omega|$ is the number of micro-states in set $\Omega$.   Since for every micro-state $\sv$, there is alway one spin-reversed state $- \sv$, $\Omega(E_0, M) = \Omega(E_0, -M)$ must be an even function of $M$.  Note that in equilibrium, all micro-states inside $\Omega(E_0, M)$ have the same energy $E_0 - B M$, and hence the same probability $e^{\beta F_{\rm EQ} - \beta (E_0 - B M)}$.  In fact, $\Omega(E_0, M)$ is a refining of the micro-canonical ensemble.  In thermal equilibrium, the total probability of Boltzmann state $\Omega(E_0, M) $ is then
\ba
p_{\rm EQ}(E_0,M) 
&=& e^{\beta ( F_{\rm EQ} - F_B(E_0,M))},
\label{p-E_0-M} \\
F_B(E_0,M) &=& E_0 - B M - T S_B(E_0, M)
\label{F_B-E_0-M}
\ea
where $F_B(E_0,M)$ shall be called the {\em Boltzmann free energy}.  


 

We note that $F_B, E_0, M$ are all extensive quantities.   In the thermodynamic limit (number of spins $N \rightarrow \infty$), the distribution Eq.~(\ref{p-E_0-M}) becomes more and more concentrated near the minimum of $F_B(E_0,M)$.  Except at a critical point,  fluctuations of all intensive quantities, such as energy and Boltzmann entropy per spin, scale as $N^{-1/2}$, i.e., they become non-stochastic in the thermodynamic limit.  This is in fact how thermodynamics emerges from the stochastic description of statistical mechanics.  {\em Minimization of Boltzmann free energy} $F_B(E_0,M) $ determines  all thermodynamic properties.  In particular, the stationarity condition reads
\begin{subequations}
\label{conditions-EQ}
 \ba
\delta^1 F_B =0  \rightarrow \left\{
\begin{array}{ll} 1 - T \frac{\partial S_B}{\partial E_0}  = 0; \vspace{3mm}\\
 - (B +  \frac{\partial S_B}{\partial M} ) = 0. 
\end{array}
\right.
 \ea
 which are in fact the {\em equations of state}.
The stability condition of the thermodynamic state reads
\be
\delta^2 F_B = - T \delta^2 S_B(E_0,M) \geq 0,
\label{stability-static}
\ee
unfolding of which gives positivity of specific heat and magnetic susceptibility. 

It is important to note that minimization of Boltzmann free energy can be understood as an application of the principle of minimal free energy in the subspace of all Boltzmann macro-states.  Indeed using the equilibrium-probability restricted inside the Boltzmann state $\Omega(E_0, M)$ in Eq.~(\ref{free-energy-def}), we obtain the Boltzmann free energy Eq.~(\ref{F_B-E_0-M}).  In the thermodynamic limit, the Boltzmann free energy and the canonical free energy must be asymptotically equivalent:
\be
F_B(\bar{E}_0, \bar{M}) = F_{\rm EQ},
\ee
\end{subequations}
otherwise the probability distribution Eq.~(\ref{p-E_0-M}) would not  be normalizable as $N \rightarrow \infty$.  Similarly Boltzmann entropy also become equivalent to the Gibbs entropy.  The underlying physics is {\em the equivalence of different statistical ensembles in the thermodynamic limit}.  


The concepts of Boltzmann macro-states and Boltzmann entropy are very flexible.  If we wish to study another macroscopic quantity $A(\sv)$, we can refine the Boltzmann state Eq.~(\ref{Boltzmann-EQ-1}) by fixing $E_0, M$ and $A$.   Then Eq.~(\ref{p-E_0-M}) becomes a joint pdf for $E_0, M$ and $A$, which can be deemed as a straightforward generalization of Einstein's theory for thermodynamic fluctuations.

\vspace{3mm}
\paragraph{Principle of minimal free action.}  We shall now  construct an ensemble theory for discrete time Markov chain dynamics with discrete-valued state variables.   Generalization of our theory to continuous time Markov processes and non-Markov processes has technical but not conceptual difficulty.  We will use $N$ to denote the number of time steps of dynamic paths.  We will take the limit $N \rightarrow \infty$, which is the dynamic analogue of thermodynamic limit.  We will not assume that the system is large, hence our theory will be applicable both for large and small systems.   Let $X_k$ be the discrete-valued state variables of the system at time step $k$. A dynamic path is described by an ordered sequence  $\gamma_N = X_N \ldots X_1X_0$, where time propagates from right to left.  Let   $p_0(X_0)$ be the initial probability distribution, and $P(X|Y)$  the transition probability.  The probability of a path $\gamma_N$ assigned by the Markov chain dynamics is 
\be
p(\gamma_N) = P(X_N|X_{N-1}) \cdots P(X_1 |X_0) p_0(X_0), 
\label{path-prob}
\ee
We define the {\em action} of a dynamic path $\gamma_N$ as 
\ba
&& {\mathcal E} (\gamma_N) 
 = - \log p(\gamma_N)
 \nonumber\\
 &=& - \sum_{k = 1}^N\log P(X_k| X_{k-1}) 
 - \log p_0(X_0).
 \label{action-def}
\ea 

To formulate an action principle for Markov chain dynamics, it is valuable to study a generic path probability distribution, denoted by $q(\gamma_N)$, that is different from Eq.~(\ref{path-prob}).   We  define the {dynamic entropy} and  the {\em free action} of a path pdf $q(\gamma_N)$ as:
\ba
H[q ] &=& - \sum_{\gamma_N} q  (\gamma_N) \log q  (\gamma_N), 
\label{dyn-entropy-def}
\\
\Phi[q] &\equiv& 
{\sum_{\gamma_N}} \,\, q(\gamma_N) 
 {\mathcal E} (\gamma_N)  - H[q  (\gamma_N)  ] 
  = D(q||p),
\label{free-action} 
\ea
where $D(q||p) \equiv \sum_{\gamma_N} q(\gamma_N) \log q(\gamma_N)  /p(\gamma_N) $ is the relative entropy, which  is known to be nonnegative and vanishes only for $q(\gamma_N)  = p(\gamma_N) $.   Just as the free energy is minimized by the equilibrium Gibbs-Boltzmann distribution, the free action Eq.~(\ref{free-action}) is minimized by $q = p$ as defined in Eq.~(\ref{path-prob}).  The proofs are exactly the same in the equilibrium and non-equilibrium cases. This is the {\bf Principle of Minimal Free Action}, which is mathematically equivalent to the definition of Markov chain dynamics.   $ {\mathcal E} (\gamma_N) $ and $H[q ]$ are the dynamic analogues of Hamiltonian and Gibbs entropy, whereas $\Phi[q] $ is the analogue of free energy.  The dynamic analogue of temperature is unity.  




\paragraph{Dynamic ensemble theory for NESS.} Our main interest in this work is non-equilibrium steady states (NESS) of time-homogeneous ergodic Markov chains.  For this purpose, it is convenient to study cyclic dynamic paths such that $X_{N} = X_0$.   We shall call such a path a loop $\ell_N = \ell (X_N\cdots X_1)$, and {\em define}  its  action $ {\mathcal E}(\ell_N)$ as
\be
{\mathcal E}(\ell_N)
=   - \log P( X_N | X_{N-1} ) 
\cdots 
P( X_1 | X_{N} ).
\label{p_loop-def}
\ee 
Using the famous Perron-Fronenius theorem, we can prove the following limit:
\be
\sum_{X_2, \ldots, X_N}  e^{- {\mathcal E}(\ell_N)}
 \rightarrow p_{\rm SS}(X_1),
\quad {N \rightarrow \infty}, 
\label{normalization-ell-1}
\ee
where $p_{\rm SS}(X_1)$ is the steady state pdf.  
Further summing Eq.~(\ref{normalization-ell-1}) over $X_1$, we obtain unity.  More specifically we have the following identity:
\be
\lim_{N \rightarrow \infty} \sum_{\ell_N} e^{- {\mathcal E}(\ell_N)} = 1. 
\label{prob-normalization}
\ee
If we use Eq.~(\ref{p_loop-def}) for the action in Eq.~(\ref{free-action}), and restrict the summation in Eq.~(\ref{free-action}) to all loops, we see that the principle of minimal free action still holds: the free action Eq.~(\ref{free-action}) is minimized by the pdf $p(\ell_N) = e^{- {\mathcal E}(\ell_N)}$.  
Restriction to loops is very convenient for the study of NESS, as we no longer have to worry about initial conditions. 


The time-reversal of a loop  $\ell_N = \ell (X_N\cdots X_1)$ is defined as $\ell^* = (X_1^*\cdots X_N^*)$, where $X^*$ is the time-reversal of state $X$. Let us define the {\em symmetric action} $\Theta(\ell)$ and {\em antisymmetric action} $\Sigma({\ell}) $ of a loop $\ell$ as:
\begin{subequations}
\label{action-decomp}
\ba
\Theta(\ell)&\equiv&
  \frac{1}{2}  {\mathcal E}(\ell_N) 
  + \frac{1}{2}  {\mathcal E}(\ell_N^*)
= \Theta(\ell^*), \\
\Sigma(\ell) &\equiv& 
 -  {\mathcal E}(\ell_N) +  {\mathcal E}(\ell_N^*)
= - \Sigma({\ell^*}) , 
\label{joint-SHT-1} \\
{\mathcal E}(\ell) &=&
 \Theta(\ell) - \frac{1}{2} \Sigma(\ell). 
\ea
\end{subequations}
 $\Theta(\ell)$ and $\Sigma({\ell}) $ are, respectively, even and odd under time reversal. 
  Using  the Detailed Fluctuation Theorem~\cite{Crooks-FT-1,Jarzynski-FT-1}, it can be easily shown that $\Sigma(\ell)$ is the total {\em entropy production} of the loop $\ell$.   In the absence of  driving force, $p(\ell)$ must be reversible, and  hence $\Sigma(\ell)$ vanishes identically.  The resulting Markov chain is then reversible. 

We shall define a {\em dynamic Boltzmann entropy}:
\be
H_B(\Theta, \Sigma) \equiv
\log | \{\ell|  \,\, \Theta(\ell) 
= \Theta, \Sigma(\ell) =\Sigma \} |. 
\label{Boltzmann-NEQ-1}
\ee
Here $|\Omega|$ is understood as the number of paths in the set $\Omega$~\footnote{To be precise mathematically, we should have used different notations for the functions $\Theta(\ell), \Sigma(\ell)$ and for the parameters $\Theta, \Sigma$, just as what we have down in equilibrium theory, see Eq.~(\ref{Boltzmann-def-static}).  To save notations, we use the same symbols.  Physicists should have no trouble distinguishing them in mind!}.   Since every loop $\ell$ has a time-reversal counterpart $\ell^*$, $H_B(\Theta, \Sigma) $ is an even function of $\Sigma$: 
$H_B(\Theta, \Sigma) = H_B(\Theta, - \Sigma)$.  We can now rewrite Eq.~(\ref{prob-normalization}) in terms of dynamic Boltzmann entropy:
\be
\sum_{\Theta, \Sigma} e^{H_B(\Theta, \Sigma)-\Theta +  \Sigma/2 }  = 1.  
\ee
The pdf of $(\Theta, \Sigma)$ is then given by
\be
P(\Theta, \Sigma) = 
e^{ - \Phi_B(\Theta, \Sigma)} =
 e^{H_B(\Theta, \Sigma) 
-\Theta +  \Sigma/2}. 
\label{F-e-sigma-1}
\ee
Here $\Phi_B(\Theta, \Sigma)$ is the dynamic analogue of Boltzmann free energy, and will be  called  {\em Boltzmann free action}: 
\be
\Phi_B(\Theta, \Sigma) \equiv - H_B(\Theta, \Sigma)  +\Theta - \frac{1}{2} \Sigma,  
\label{Phi_B-def}
\ee 
which completely determines the probability distribution Eq.~(\ref{F-e-sigma-1}).  Because $H_B(\Theta, \Sigma)$ is even in $\Sigma$, we easily see that $P(\Theta, \Sigma) = P(\Theta, - \Sigma) e^{\Sigma}$.  Summing this relation over $\Theta$, we obtain the famous Steady-State Fluctuation Theorem:
$P(\Sigma) = P( - \Sigma) \, e^{\Sigma}$.  

Note that $H_B, \Theta, \Sigma$ are all extensive in number of time steps $N$.  As $N$ becomes large, $P(\Theta, \Sigma) $ becomes more and more concentrated near the minimum of $\Phi_B(\Theta, \Sigma)$, with a width scaling sub-extensively as $\sqrt{N}$.  The $N \rightarrow \infty$ limit is clearly the  dynamic counterpart of thermodynamic limit.  Mathematicians call this {\em large deviation limit}, and would demand a mathematical proof for its existence.  We shall argue that its existence is intuitive.   The most probable values $(\bar{\Theta}, \bar{\Sigma})$ are  determined by
minimization of the Boltzmann free action $\Phi_B$:
\begin{subequations}
\label{saddle-point-1}
\ba
\delta^1 \Phi_B = \delta H_B - \delta \Theta + \frac{1}{2} \delta \Sigma &=& 0,
\label{stationarity-dyn-1} \\
\delta^2 \Phi_B =\delta^2 H_B &\leq& 0, 
\label{stability-dyn-1} \\
 H_B(\bar{\Theta}, \bar{\Sigma}) 
 -\bar{\Theta}+ \frac{1}{2} \bar{\Sigma} &=& 0. 
\label{normalization-1}
\ea
\end{subequations} 
These results are the dynamic analogues of Eqs.~(\ref{conditions-EQ}), and are valid only in the sense of extensive variables in $N$.  Specifically Eq.~(\ref{stationarity-dyn-1})  determines the equations of state for NESS, (\ref{stability-dyn-1}) is the condition of stability, and also determines the small fluctuations of $\Theta, \Sigma$.  Finally Eq.~(\ref{normalization-1}) is demanded by normalization of probability, and signifies the equivalence of different dynamic ensembles.   All these identities have analogues in equilibrium ensemble theory.  

\vspace{2mm}
\paragraph{A simple example.} We discuss a simple example to illustrate the dynamic ensemble theory.  Consider a particle hopping randomly on a circle.  An external force is applied so that the particle hops asymmetrically.  The probabilities that the particle hop clockwise or counter-clockwise, or idle, are respectively $z(\delta)^{-1} \,e^{- \varepsilon_0 + \delta}, z(\delta)^{-1} \,e^{- \varepsilon_0 -\delta}$ and $ z(\delta)^{-1} $, where $\delta$ is the driving force, and $z(\delta) = 1+ 2 e^{-\varepsilon_0  } \cosh\delta$ is a normalization constant.   Let $N_{\pm}(\ell), N_0(\ell)$ be the numbers of CW, CCW, and idle steps in a loop $\ell$,  and $p_{\pm}(\ell) = N_{\pm}(\ell)/N, p_0(\ell) = N_0(\ell)/N$ are the empirical frequencies.  $p_{\pm} , p_0$ are the macroscopic variables we use to characterize each loop, which satisfy $p_+ + p_- + p_0 = 1$.   The action ${\mathcal E}(\ell) $ is
\ba
&& {\mathcal E}(\ell) = - \log p(\ell)\\
&=& N \left( \log z(\delta) + (p_+(\ell) + p_-(\ell) ) \varepsilon_0 
- (p_+(\ell) - p_-(\ell)) \delta \right). 
\nonumber
\ea
Comparing this with Eq.~(\ref{action-decomp}) we see that  the entropy production of the loop $\ell$ is $ \Sigma(\ell) =2 N  (p_+(\ell) - p_-(\ell)) \delta $. 
The dynamic Boltzmann entropy as a function of macroscopic variables $p_{\pm}, p_0$ can also be easily calculated:
\ba
H_B &=& \log \frac{N!}{N_0 ! N_+! N_-!} 
\\
&=& - N \left( p_0 \log p_0 + p_+ \log p_+ + p_- \log p_-
\right) .  \nonumber
\ea
Substituting these back into Eq.~(\ref{Phi_B-def}) and minimizing, we find the mean entropy production:
 \be
\bar{ \sigma } = 2 (p_+ - p_-) \delta 
= \frac{4 \, e^{- \varepsilon_0}\delta  \sinh \delta}
{ 1+ 2 \, e^{- \varepsilon_0}  \cosh \delta}.
 \ee
Stability of the NESS can be easily verified by computing the second order derivative of $\Phi_B(p_+,p_-)$.  We note that this example can be solved using elementary method.  Here we use it only to illustrate the structure of dynamic ensemble theory.  In future publications, we will use the dynamic ensemble theory to study more realistic many-body non-equilibrium problems.  


\vspace{2mm}
\paragraph{Quadratic Approximation.} Consider a system driven by two weak external forces $\lambda_1, \lambda_2$.  We can expand the antisymmetric action $\Sigma(\ell)$ in terms of $\lambda_{1,2}$: 
\be
\Sigma(\ell) =2  \lambda_1 Q_1(\ell) 
+ 2  \lambda_2 Q_2(\ell) + O(\lambda^3), 
\label{Sigma-ell-def-1}
\ee
where $Q_1(\ell), Q_2(\ell)$ are called dissipative currents, and are odd under time-reversal, $Q_{1,2}(\ell^*) = - Q_{1,2}(\ell)$.  The symmetric action $\Theta (\ell)$  also depends on $\lambda_{\alpha}$, via the normalization condition $\sum_{\ell} \exp ( - \Theta (\ell) + \Sigma(\ell)/2) = 1$.  A simple calculation shows that   (with Einstein's summation convention used)
\ba
\Theta(\ell) &=& \Theta_0(\ell) -  
\log \left\langle e^{ \lambda_{\alpha}Q_{\alpha}(\ell)} 
\right\rangle_0
\nonumber\\
&=&  \Theta_0(\ell) -  \frac{1}{2} \lambda_{\alpha} L_{\alpha \beta} \lambda_{\beta}
 + O(\lambda^3),
\ea
where $ L_{\alpha \beta}  = \langle Q_{\alpha} Q_{\beta} \rangle_0$ is the correlation function of dissipative currents in the absence of $\lambda_{1,2}$.  In the last step, we have expanded in terms of $\lambda$, and have used the fact that $\langle Q_{\alpha} \rangle_0 \equiv 0$, i.e., time-reversal symmetry in the equilibrium case. 

It is most convenient to treat the dynamic Boltzmann entropy as a function of $\Theta_0$ and $Q_{\alpha}$. As such, $H_B(\Theta_0, Q)$ is independent of $\lambda_{\alpha}$.  Furthermore, because of time-reversal symmetry, $H_B(\Theta_0, Q)$ must be even in $Q$.
%
We expand $H_B$ and $\Phi_B$ up to $Q^2$:
\ba
&& H_B(\Theta_0,Q) = H_B(\Theta_0,0) 
-  \frac{1}{2} Q_{\alpha}  {L^{-1}}_{\alpha\beta} Q_{\beta},\\
&& \Phi_B(\Theta_0,Q)  
=  \Phi_B(\Theta_0,0)
+ \frac{1}{2} (Q - L  \lambda)_{\alpha} L^{-1}_{\alpha \beta}
 (Q - L  \lambda)_{\beta}. 
\nonumber
\ea
As such $ H_B(\Theta_0,Q)$ becomes identical to the dissipation functional defined by Onsager \cite{Onsager-1931,Onsager-Machlup-1953}, if we identify $\lambda_{\alpha}$ with generalized forces.   The average currents are determined by minimization of  $\Phi_B(\Theta_0,Q)$, which yield result $\bar Q_{\alpha} = L_{\alpha \beta} \lambda_{\beta}$.   The reciprocal relations are already encoded by the symmetry current correlations $L_{\alpha \beta} = L_{\beta \alpha}$.  It is however important to note that $ \Phi_B(\Theta_0,Q) $ does not describe dissipation of energy, or production of entropy.  Also it would be completely wrong to call Eq.~(\ref{saddle-point-1}) the {\em principle of least dissipation of energy}, or {\em the principle of minimal entropy production.}   As we have shown through out this work, the entropy production changes sign under time-reversal, has no upper bound or lower bound, and certainly can not achieve minimum or maximum at the steady state.  In fact, as we have demonstrated, it is the principle of minimal free action that governs the physics of non-equilibrium steady states. 

\vspace{2mm}
\paragraph{From minimal free action to minimal free energy.}  Similar to the equilibrium ensemble theory, the dynamic ensemble theory can be used to study arbitrary macro-variables.  All we need to do is to refine the definition of dynamic Boltzmann entropy Eq.~(\ref{Boltzmann-NEQ-1}) by specifying the additional macro-variable $A(\ell)$  that we aim to study: 
\be
H_B(\Theta, \Sigma, A) \equiv
\log | \{\ell|  \,\, \Theta(\ell) 
= \Theta, \Sigma(\ell) =\Sigma, A(\ell) = A \} |.
\nonumber
\ee

As an example, we may study the empirical pair  distribution  of a loop $\ell$:
\be
f(x,y;\ell) = \frac{1}{N} \sum_{k=1}^N \delta(x, X_k) \delta (y, X_{k+1}),
\ee
which gives the frequency of transition from state $x$ to state $y$ during the dynamic path $\ell_N$.   It is normalized as $\sum_{x,y} f(x,y; \ell)  = 1$, and hence is an intensive quantity. The symmetric and antisymmetric actions of the loop can be calculated in terms of $f(x,y;\ell)$:
\begin{subequations}
\ba
\Theta (\ell) &=& - \frac{1}{2} \sum_{x,y} f(x,y;\ell) \log \left( P(y|x) P(x|y) \right),\\
\Sigma(\ell) &=&  \sum_{x,y} f(x,y;\ell) \log \frac{P(y|x)}{P(x|y)}.
\label{Sigma-f-1} 
\ea
\end{subequations}
The {\em empirical state distribution}, which gives the frequency of state $x$ appearing in the path $\ell$, can also be calculated in terms of $f(x,y;\ell)$:
\be
p(x;\ell) = \frac{1}{N} \sum_{y} f(x,y;\ell)
 = \frac{1}{N} \sum_{y} f(y,x;\ell). 
\ee
We can similarly define the dynamic Boltzmann entropy and free action as functional of $f(x,y;\ell)$, and construct the pdf of $f(x,y)$:
\begin{subequations}
\ba
H_B[f]  &\equiv&
\log |\{ \ell| f(x,y;\ell) = { f}(x,y) \}|,  \\
\Phi_B[f] &\equiv& - H_B[f] +\Theta[f] -  \Sigma[f]/2,\\
P[f] &=& \exp -\Phi_B[f]. \label{P-f-1}
\ea
\end{subequations}
 In fact, $\Phi_B[f]$ can be calculated using results of large deviation theory~\footnote{For derivation of this result, see reference~\cite{Large-deviation-Dembo}. $\Phi_B(f)$ is called level-3 large deviation rate function of the Markov chain. }:
\ba
\Phi_B[f] 
= N  \sum_{x,y} f(x,y) \log \frac{f(x,y)}{P(y|x) f_1(x)},
\ea
where $f_1(x) = \sum_y f(x,y) = \sum_y f(y,x)$ is the marginal distribution of $f(x,y)$.  Again $H_B[f] $ is extensive in $N$, and distribution of $f(x,y)$ becomes concentrated near the saddle point $\bar{f}(x,y)$, which must satisfy $\bar{f}(x,y) = P(y|x)\bar{f}_1(x)$.  But this precisely means that $\bar{f}(x,y)$ is the stationary pair distribution, and $\bar{f}_1(x) = p_{SS}(x)$ is the stationary state distribution.  

Note that  $\Theta(\ell), \Sigma(\ell)$, $f(x,y;\ell), f_1(x,\ell)$ are all properties of individual dynamic path.  Mathematically Eqs.~(\ref{F-e-sigma-1}) and (\ref{P-f-1}) say that these  macro-variables {\em converge to their means in probability}.  A dynamic path is called {\em typical} if it minimizes the Boltzmann free action.  Typical paths are selected by the principle of minimal free action.  Our results then say that a typical dynamic path exhibits average entropy production $\bar{\Sigma}$, average symmetric action $\bar{\Theta}$, and average distribution of transition frequency $\bar{f}(X,Y)$, as well as average distribution of states $p_{SS}(X)$.  For time-reversal Markov chains, we know that $p_{SS}(X)$ is the Gibbs-Boltzmann distribution, which minimize the free energy.   This means that if we pick a long, typical dynamic path, and construct its empirical state distribution, we will precisely find the Gibbs-Boltzmann distribution--This is ergodicity at work in a Markov chain model! In another word,  {\em the principle of minimal free action gives a dynamic justification of the principle of minimal free energy. }  

\vspace{3mm}
This work is supported by NSFC via grant \#11674217, as well as Shanghai Municipal Education Commission and Shanghai Education Development Foundation via ``Shu Guang'' project.





\end{document}